\documentclass[12pt]{article}
\usepackage{amsfonts}
\usepackage[dvips]{graphics}
\newcommand{\Rl}{\mathbb{R}}

\newcommand{\Ir}{\mathbb{Z}}
\newcommand{\Cx}{\mathbb{C}}

\newcommand{\HH}{\mathcal{H}}

\newtheorem{theorem}{Theorem}[section]
          
\newtheorem{lemma}[theorem]{Lemma}

\newcommand{\rem}[1]{{\bf Remark:}}
\newcommand{\condmat}[1]{archived as {\tt cond-mat/#1}}
\newcommand{\Section}[1]{\setcounter{equation}{0}\section{#1}}

\newcommand{\eq}[1]{(\ref{#1})}
\newenvironment{proof}{\noindent {\bf Proof: }}{\QED\medskip}
\def\QED{{\hspace*{\fill}{\vrule height .5ex width 1ex }\quad} 
    \vskip 0pt plus20pt}
\newcommand{\be}{\begin{equation}}
\newcommand{\ee}{\end{equation}}
\newcommand{\bea}{\begin{eqnarray}}
\newcommand{\eea}{\end{eqnarray}}
\newcommand{\beann}{\begin{eqnarray*}}
\newcommand{\eeann}{\end{eqnarray*}}

\newcommand{\ket}[1]{\left\vert #1\right\rangle}
\newcommand{\bra}[1]{\left\langle #1\right\vert}

\DeclareMathAlphabet{\mathol}{OT1}{cmr}{l}{ol}

\newcommand{\up}{\ket{\uparrow}}
\newcommand{\down}{\ket{\downarrow}}

\newcommand{\ud}{\ket{\uparrow \downarrow}}
\newcommand{\du}{\ket{\downarrow \uparrow}}

\newcommand{\C}{\mathbb{C}}

\newcommand{\ip}[2]{\langle{#1|#2}\rangle}
\newcommand{\op}[3]{\bra{#1} #2 \ket{#3}}

\newcommand{\unity}{1\hskip -3pt \rm{I}}

\newcommand{\mod}{\, {\rm mod}\, }
\newcommand{\real}{\, {\rm Re}}
\newcommand{\sech}{\, {\rm sech}}

\begin{document}
{\baselineskip=10pt \thispagestyle{empty} {{\small Preprint UC Davis Math
1999-30}
\hspace{\fill}}

\vspace{20pt}

\begin{center}
{\LARGE \bf A continuum approximation for the excitations
of the $(1,1,\dots,1)$\\
interface in the quantum Heisenberg model\\[27pt]}
{\large \bf Oscar Bolina, Pierluigi Contucci, Bruno Nachtergaele 
and Shannon Starr\\[10pt]}
{\large  Department of Mathematics\\
University of California, Davis\\
Davis, CA 95616-8633, USA\\[15pt]}
{\normalsize bolina@math.ucdavis.edu, contucci@math.ucdavis.edu, 
bxn@math.ucdavis.edu, sstarr@math.ucdavis.edu}\\[30pt]
\end{center}

\noindent
{\bf Abstract:}
It is shown that, with an appropriate scaling, the energy of low-lying 
excitations of the $(1,1,\dots,1)$ interface in the $d$-dimensional
quantum Heisenberg model are given by the spectrum of the $d-1$-dimensional 
Laplacian on an suitable domain.

\vspace{8pt}
\vskip .2 cm
\noindent
{\small \bf MCS2000 numbers:} 82B10, 82B24, 82D40 
\vfill
\hrule width2truein \smallskip {\baselineskip=10pt \noindent Copyright
\copyright\ 1999 Bolina, Contucci, Nachtergaele, and Starr. Reproduction of 
this article in its entirety, by any means, is permitted for non-commercial 
purposes.\par }}

\newpage

\Section{Introduction and main results}

We consider the spin 1/2 XXZ Heisenberg model on the $d$-dimensional
lattice $\Ir^d$. For any finite volume $\Lambda\subset\Ir^d$, the 
Hamiltonian is given by 
\be
H_\Lambda = - \sum_{x,y\in\Lambda\atop \vert x-y\vert=1}
\Delta^{-1} (S_x^{(1)} S_y^{(1)} + S_x^{(2)} S_y^{(2)}) 
+ S_x^{(3)} S_y^{(3)},
\ee
where $\Delta>1$ is the anisotropy. We refer to the next section for more
precise definitions. By adding an appropriate boundary term one can 
insure that the ground states of this model describe an interface
in the $(1,1,\dots,1)$ direction between two domains with opposite magnetization.
For a particular choice of boundary term, the model has exactly one
ground state $\psi_n$ for each fixed number of down spins, $n$. We call these
the canonical ground states. In analogy with statistical mechanics of
particle systems one can introduce the grand canonical ground states
of the form
$$
\Psi =\sum_n z^n \psi_n
$$
It turns out that these states are inhomogeneous product states
\cite{GW}. In this paper, we consider a class of perturbations of these
product states, of which we calculate the energy. By the variational 
principle this leads to bounds for the energy of the first excited state
of the model. As the excitation spectrum above the interface states
is gapless \cite{KN2, Mat2}, this bound should vanish as the volume 
tends to infinity. This is indeed the case (see \eq{energy:norm}).

The perturbations we consider are in correspondence with functions 
$f : \Lambda \to \Cx$.
Furthermore, we consider functions which are slowly-varying
in all directions perpendicular to $(1,1,\dots,1)$ though they may have 
discrete jumps parallel to this direction.
In other words $\|\nabla f \cdot v\|_\infty \ll \|f\|_\infty$ for all
$v \perp (1,1,\dots,1)$.
We consider general perturbations of this type and
conclude that the optimal perturbations, in the sense of minimizing energy,
are localized near the interface.
With this restriction, the Hamiltonian, projected to and restricted to the 
appropriate subspace, is just the Laplacian

This result may be compared to the recent bound of
\cite{BCNS}.
The main dif\-ference is that there we considered a canonical ensemble,
for which there were a fixed number of down-spins (hence a fixed number
of up-spins).
We developed a version of equivalence of ensembles whereby we
estimated the canonical expectation of a gauge invariant observable
by a grand canonical expectation, provided that the interfaces of the 
canonical and grand canonical states occupied the same position.

In the present paper, we begin with the grand canonical ensemble, 
so that we make no reference to equivalence of ensembles.
Specifically, we consider a cylindrical region of total height $L+1$
and whose cross-section is a region $\Omega_R$ with linear size $R$.
Then a class of excitations is parametrized by smooth functions $\Phi$
on a fixed domain $\Omega = R^{-1} \Omega_R$.

{\bf Main Result:}
{\em Excitations on $\Lambda$ have a normalized energy
\be
\label{energy:norm}
\frac{\op{\psi^f}{H}{\psi^f}}{\ip{\psi^f}{\psi^f}}
  \approx\frac{1}{2 \Delta R^2} \cdot 
  \frac{\|\nabla\Phi\|^2_{L^2(\Omega)}}{\|\Phi\|^2_{L^2(\Omega)}}
  \cdot g(\Delta, \mu)
\ee
where 
$$
g(\Delta,\mu)=\frac{\sum_{l=-L/2}^{L/2-1} \sech(\alpha[l-\mu]) 
  \sech(\alpha[l+1-\mu])}{\sum_{l=-L/2}^{L/2} \sech(\alpha[l-\mu]) 
  \sech(\alpha[l-\mu])}.
$$
Here, $\mu$ is a real parameter of the grand canonical ground
state describing the location of the interface between the regions 
of homogeneous up and down spins.
As $\mu \to -\infty$, the ground state has all spins up, 
and for $\mu \to \infty$, all spins are down.
For all $\mu\in \Rl$, and sufficiently large $L$, $g$ satisfies the bounds
$$
\frac{1}{2\Delta}\leq \Delta-\sqrt{\Delta^2-1}\leq g(\Delta,\mu)\leq 1
$$}

\begin{rem}{}
The normalized energy of \eq{energy:norm}
is exactly the same as that for the Laplacian.
Equating the first variation to zero, we see that
the local extrema of the normalized energy are precisely the solutions of
$\nabla^2 \Phi = - \lambda \Phi$ (here $\nabla^2$ is the Laplacian),
and $\lambda = \|\nabla\Phi\|^2_{L^2(\Omega)}/\|\Phi\|^2_{L^2(\Omega)}$.
The space of excitations we consider does not form an invariant subspace
of $H$,
so that the eigenvectors of the Laplacian are not truly eigenvectors of $H$.
But, using the variational inequality, we see that the spectral gap of
$H$ is bounded thus:
$$
\gamma_1 \leq 
  \frac{\lambda_1}{2 \Delta R^2} \cdot g(\Delta, \mu) (1 + O(\frac{1}{R^2})),
$$
where $\lambda_1$ is the first positive eigenvalue of $-\nabla^2$ with
Dirichlet boundary conditions on the domain $\Omega$.
\end{rem}

\Section{The Spin-$\frac{1}{2}$ Heisenberg XXZ Ferromagnet}

A quantum spin model, such as the Heisenberg XXZ ferromagnet,
is defined in terms of a family of local Hamiltonians $H_\Lambda$, acting
as self-adjoint linear operators on a Hilbert space $\HH_\Lambda$. This
family is parametrized by finite subsets $\Lambda\subset \Ir^d$. 

We choose $\Lambda$ to be ``cylindrical'' in the following sense:
Let $\{e_j\}_{j=1}^d$ be the set of coordinate unit vectors
and define the vector $e_* = \sum_{j=1}^d e_j = (1,1,\dots,1)$,
which is the axial direction for the cylinder.
Define the functional  $l(x) = x \cdot e_* = \sum_{j=1}^d x^j$,
where $x = \sum_{j=1}^d x^j e_j$.
Observe that the kernel of $l$ in $\Ir^3$ is a $(d-1)$-dimensional 
sublattice perpendicular to the axial direction.
Take for the base of $\Lambda$ a finite subset of this $(d-1)$-dimensional 
sublattice, and call it $\Gamma$. 
A discrete approximation to the line of all scalar mutliples of $e_*$
is the \textit{one-dimensional stick} $\Sigma$.
$\Sigma$ is a bi-infinite sequence of points $\{x_n\}_{n=-\infty}^\infty$
such that $x_0=0$ and all other points $x_n$ are specified by the relation
$x_n - x_{n-1} = e_{n \mod d}$.
So 
\begin{eqnarray*}
\Sigma &=& \{\ldots,-(e_d+e_{d-1}+\cdots+e_1+e_d), -(e_d+e_{d-1}+\cdots+e_1), 
  \ldots, -e_d, \\
  &&  0, e_1, (e_1+e_2), \ldots,
  (e_1+e_2+\cdots+e_d), (e_1+e_2+\cdots+e_d+e_1), \ldots\}.
\end{eqnarray*}
A finite stick of length $L+1$, where $L$ is even, is 
$\Sigma_{L} = \{x \in \Sigma : -L/2 \leq l(x) \leq L/2\}$.
Now define $\Lambda$ to be the translates of $\Gamma$ along $\Sigma_{L}$,
i.e.\ 
\be
\Lambda = \Gamma + \Sigma_L = \{x + y : x\in \Gamma, y \in \Sigma_L\}.
\ee

Let us now define \textit{nearest neighbors} to be points $x,y \in \Ir^d$
such that $|l(x) -  l(y)|=1$ and $\|x - y\|_{l^1} = 1$.
Also, we define oriented bonds between nearest neighbors as 
ordered pairs $(x,y)$ satisfying $l(y) = l(x) + 1$
and $\|x - y\|_{l^1} = 1$.
Hence $\{(x,x + e_j)\}_{j=1}^d$ is the set of all
oriented bonds with lower point $x$. 
The collection of all oriented bonds with both points in $\Lambda$,
will be called $B(\Lambda)$.

The local Hilbert spaces are $\HH_\Lambda = (\Cx^2)^{\otimes |\Lambda|}$.
Each copy of $\Cx^2$ comes with an ordered basis $(\up,\down)$
and a spin-$\frac{1}{2}$ representation
of $SU(2)$ defined by the Pauli matrices:
\be
S^{(1)} = \left(\begin{array}{cc} 0 & 1/2 \\
1/2 & 0 \end{array}\right),\quad
S^{(2)} = \left(\begin{array}{cc} 0 & -i/2 \\
i/2 & 0 \end{array}\right),\quad
S^{(3)} = \left(\begin{array}{cc} 1/2 & 0 \\
0 & -1/2 \end{array}\right).		
\ee
(So, for example, $S^{(3)} \up = \frac{1}{2} \up$ and
$S^{(3)} \down = -\frac{1}{2} \down$.)
We consider a family of Hamiltonians parametrized by a real
number $\Delta \geq 1$.
In order to define the total Hamiltonian, we first define 
local Hamiltonians $h_{x y}$ for each oriented bond $(x,y)$:
\be
h_{x,y} = - \Delta^{-1} (S_{x}^{(1)} S_{y}^{(1)}
+ S_{x}^{(2)} S_{y}^{(2)}) - S_{x}^{(3)} S_{y}^{(3)} 
+ \frac{1}{4} + \frac{1}{4} A(\Delta) (S_{y}^{(3)} - S_{x}^{(3)}),
\ee
where $A(\Delta) = {1 \over 2} \sqrt{1 - 1/\Delta^2}$.
The total Hamiltonian is
\be\label{ham}
H_\Lambda = \sum_{(x,y)\in B(\Lambda)} h^{q}_{x,y}.
\ee
$\Delta$ parametrizes ``anisotropic coupling''.
The case $\Delta = 1$ is the isotropic model, also known as the 
Heisenberg XXX ferromagnet, which exhibits $SU(2)$ symmetry 
(because $H_\Lambda$ commutes with $S^1$, $S^2$ and $S^3$).

We find it convenient to introduce a positive constant
$\alpha$, which solves $\Delta = \cosh(\alpha)$.
We note that the nearest neighbor interaction $h_{x y}$ is an 
orthogonal projection 
\be
h_{x y} = \ket{\xi_{x y}} \bra{\xi_{x y}} 
  \otimes \unity_{\Lambda \setminus (x,y)},
\ee
where
\be
\xi_{x y} = \frac{e^{-\alpha/2} \du - e^{\alpha/2} \ud}
  {\sqrt{2 \cosh(\alpha)}}. \label{def:xi}
\ee
This also shows that each $h_{x y}$ is a  nonnegative
self-adjoint operator, hence $H_\Lambda$ is, as well.
To simplify the notation we will often drop the subscript 
$\Lambda$ when the volume is obvious from the context.

\Section{Ground States and a Perturbation}

The ground states of the XXZ ferromagnet can be calculated exactly \cite{ASW}.
We will choose a particular ground state and construct  an orthogonal subspace
(but not the entire orthogonal complement)  which is parametrized by
$H^1$-functions on a compact domain  $\Omega_0 \subset \Rl^{d-1}$. 
The inner product becomes approximately the $L^2$ inner-product and the  
orthogonal projection of the Hamiltonian is approximately the Laplacian.

The lowest eigenvalue for $H$, which is zero, has a 
$(|\Lambda|+1)$-fold degeneracy in the eigenspace.
This space of ground states is spanned by the simple tensor ground states,
which we will call grand canonical states.
Specifically, let $z$ be any complex number, and $\mu = \real(z)$.
Define the vector
\be
v_x(z) = \frac{e^{\alpha (l_x-z)/2} \up 
  + e^{-\alpha (l_x-z)/2} \down}
  {\sqrt{2 \cosh(\alpha[l_x - \mu])}} ,
\ee
for each site $x \in \Lambda$.
We define the product of these vectors
\be
\psi_0(z) = \bigotimes_{x \in \Lambda} v_x(z) ,
\ee
and we may quickly establish that it is a ground state.
Indeed, the oriented bonds are defined between points $x$ and $y$ with 
$l(y) = l(x) + 1$,
from which we see 
\be
\ip{\uparrow \downarrow}{v_{x}(z) \otimes v_{y}(z)} 
  = e^{\alpha} \ip{\downarrow \uparrow}{v_{x}(z) \otimes v_{y}(z)}.
\ee
This implies $v_{x}(z) \otimes v_{y}(z)$ is orthogonal to $\xi_{x y}$,
for each $(x,y) \in B(\Lambda)$, which proves that 
$\psi_0(z)$ is a ground state.
As we have said, the states $\psi_0(z)$ span the entire ground state
space, as $z$ ranges over all the complex numbers \cite{GW}. 
(More than this can be said.
The simple tensor ground states are parametrized by elements of 
$\Cx P^1$, so that the submanifold of all such states in $\HH$
is topologically a sphere.
But to obtain the north and south poles of the sphere, it is necessary to
take the limits $z \to \infty$ and $z \to -\infty$.) 

Let us now fix $z$, and for simplicity we will just write 
$\psi_0$ and $v_x$ without explicit reference to $z$.
For each site $x$ we define a vector orthogonal to $v_x$, 
\be
w_x = \frac{e^{-\alpha (l_x-\bar{z})/2} \up 
  - e^{\alpha (l_x-\bar{z}/2} \down}
  {\sqrt{2 \cosh(\alpha[l_x - \mu])}} .
\ee
We will make use of $w_x$ to define an orthonormal system
of states
\be
\psi^x = w_x \otimes \bigotimes_{y \in \Lambda \setminus x}
  v_y ,
\ee
where $x$ ranges over $\Lambda$.
Each of these states is also orthogonal to $\psi_0$,
let us call their span $V$. An arbitrary state in $V$ is
characterized by a function $f:\Lambda \to \Cx$.
Explicitly, $\psi^f = \sum_{x \in \Lambda} f(x) \psi^x$.
It is then clear that
$\ip{\psi^f}{\psi^g} = \sum_{x \in \Lambda} \overline{f(x)} g(x)$.

Our interest is the case that $\Lambda \nearrow \Ir^d$,
i.e.\ the thermodynamic limit.
In terms of $v_x$ and $w_x$, we see that
the local interaction $h_{x y}$ 
describes a nearest-neighbor interaction.
It may be interpreted as a bilinear form, which is a first order 
finite-difference operator in each variable.
To be clear, a straightforward calculation gives
\begin{eqnarray}
\label{local:en:exact}
\op{\psi^f}{h_{x y}}{\psi^g}
  &=& \frac{1}{2} \sech(\alpha) \sech(\alpha[l_{x}-\mu]) 
  \sech(\alpha[l_{y}-\mu]) \nonumber \\
  && \times \big(\cosh(\alpha[l_{y}-\mu]) \overline{f(y)}
  - \cosh(\alpha[l_{x}-\mu]) \overline{f(x)}\big) \nonumber \\
  && \times \big(\cosh(\alpha[l_{y}-\mu]) g(y)
  - \cosh(\alpha[l_{x}-\mu]) g(x)\big) .
\end{eqnarray}
Recall that $\mu = \real(z)$) and the energy is
\be
\label{total:en}
\op{\psi^f}{H}{\psi^g}
  = \sum_{l = -L/2}^{L/2-1}\ \sum_{x \in \Gamma_l}\ \sum_{j=1}^d
  \op{\psi^f}{h_{x,x+e_j}}{\psi^g},
\ee
where $\Gamma_l$ refers to the set of points $x \in \Lambda$
with $l(x) = l$.
In the thermodynamic limit, 
we may scale the plane $e_*^\perp = \{v \in \Rl^d : v \cdot e_* = 0\}$
so that $H$ becomes, to first order, a differential 
operator with respect to each direction of the plane.
However, the inhomogeneity in the $e_*$ direction admits no such scaling
for that coordinate, so that $H$ is genuinely a finite-difference operator
even in the thermodynamic limit.

This intuitive description of the last paragraph
is made precise, now.
Let $\Omega$ be a bounded, open subset of $e_*^\perp$
with a $C^1$ boundary.
Let $\Omega_R$ be the dilation $R \cdot \Omega = \{R x : x \in \Omega\}$,
and let $\Gamma = \Omega_R \cap \Ir^d$ be the discrete approximation
to $\Omega_R$.
As before, $\Gamma$ is the base of $\Lambda$.
Now we choose a smooth, complex-valued function $\Phi$ 
on $\Omega$, and extend it to the infinite cylinder 
$\Omega \times \Rl e_*$ so that $\nabla \Phi \cdot e_* = 0$.
(In other words, $\Phi$ is constant along the direction $e_*$.)
Let $\phi(x) = \Phi(x/R)$, which is defined on 
$\Omega_R \times \Rl e_*$ with the 
property that $\nabla \phi \cdot e_* = 0$.
Finally, let $f(x) = F(l_x) \phi(x)$, where
$F$ is a sequence $F(-L/2), \dots, F(L/2)$.
Note that $f$ is not the most general form possible
for a function on $\Lambda$, most notably because it is the product 
of functions which vary on perpendicular subspaces.
However, the span of such functions does correspond to all of 
$V$ for a fixed value of $L$ and $R$.

Next we consider the norm and energy for such a state.
We will introduce estimates for these quantities, but we will
postpone the actual error terms until the next section.
First we replace the sum over $\Gamma$ with the integral
over $\Omega$, and thus obtain an expression for the norm:
\begin{eqnarray}
\label{norm:approx}
\ip{\psi^f}{\psi^f} 
  &=& \sum_{l=-L/2}^{L/2} \ \sum_{x \in \Gamma_l} |f(x)|^2 \nonumber\\
  &\approx& |\Gamma| \ \sum_{l=-L/2}^{L/2} |F(l)|^2 \cdot
  \frac{1}{m(\Omega_R)} \int_{\Omega_R} |\phi(x)|^2\, dx \nonumber\\
  &=& |\Gamma| \ \sum_{l=-L/2}^{L/2} |F(l)|^2 \cdot
  \frac{1}{m(\Omega)} \int_{\Omega} |\Phi(x)|^2\, dx .
\end{eqnarray}
To obtain an approximation for $\op{\psi^f}{H}{\psi^f}$, we decompose
a step of $f$ along a coordinate direction into a step parallel 
to $e_*$ and a step perpendicular to $e_*$,
\begin{eqnarray*}
f(x+e_j) 
  &=& F(l_x+1) \phi(x+e_j) \\
  &\approx& F(l_x+1) \phi(x) + F(l_x+1) \nabla \phi(x) \cdot e_j.
\end{eqnarray*}
Then using the fact that
$$\sum_{j=1}^d \nabla \phi(x) \cdot e_j = \nabla \phi(x) \cdot e_* = 0,$$
and referring to \eq{local:en:exact} and \eq{total:en}
we have the apparently cumbersome expression
\begin{eqnarray*}
\op{\psi^f}{H}{\psi^f}
  &\approx& \frac{3 |\Gamma|}{2 \cosh(\alpha)} \cdot 
  \frac{1}{m(\Omega)} \int_\Omega |\Phi(x)|^2 dx \\
  && \quad \times \sum_{l=-L/2}^{L/2} 
  \Big[\sech(\alpha[l-\mu]) \sech(\alpha[l+1-\mu]) \\
  && \qquad |\cosh(\alpha[l+1-\mu]) F(l+1) 
  - \cosh(\alpha[l-\mu]) F(l)|^2\Big] \\
  && + \frac{|\Gamma|}{2 R^2 \cosh(\alpha)} 
  \cdot \frac{1}{m(\Omega)} \int_\Omega |\nabla \Phi(x)|^2 dx \\
  && \quad \times \sum_{l=-L/2}^{L/2} 
  \sech(\alpha[l-\mu]) \cosh(\alpha[l+1-\mu]) |F(l+1)|^2 .
\end{eqnarray*}

We notice that the first summand is order 1, while the second summand 
is order $1/R^2$.
We wish to minimize the energy in the limit $R \to \infty$,
so it seems sensible to eliminate the order 1 summand.
This is accomplished by letting $F(l) = \frac{1}{2} \sech(\alpha[l-\mu])$,
or any constant multiple thereof.
One point of interest is that the perturbation takes place 
primarily in a neighborhood of the interface.
The expression for the energy is
\begin{eqnarray}
\label{energy:approx}
\op{\psi^f}{H}{\psi^f}
  &\approx& \frac{|\Gamma|}{8 R^2 \cosh(\alpha)}
  \cdot \frac{1}{m(\Omega)} \int_\Omega |\nabla \Phi(x)|^2 dx \nonumber \\
  && \quad \times \sum_{l=-L/2}^{L/2-1} 
  \sech(\alpha[l-\mu]) \sech(\alpha[l+1-\mu]) .
\end{eqnarray}
Similarly, \eq{norm:approx} may be rewritten as
\begin{eqnarray}
\label{norm:approx2}
\ip{\psi^f}{\psi^f}
  &\approx& \frac{|\Gamma|}{4}
  \cdot \frac{1}{m(\Omega)} \int_\Omega |\Phi(x)|^2 dx 
  \cdot \sum_{l=-L/2}^{L/2} \sech^2(\alpha[l-\mu]) .
\end{eqnarray}
Taking the ratio, we arrive at a normalized energy
\begin{eqnarray}
\frac{\op{\psi^f}{H}{\psi^f}}{\ip{\psi^f}{\psi^f}}
  &\approx& \frac{\sech(\alpha)}{2 R^2} \cdot 
  \frac{\|\nabla\Phi\|^2_{L^2(\Omega)}}{\|\Phi\|^2_{L^2(\Omega)}} \nonumber\\
  && \times
  \frac{\sum_{l=-L/2}^{L/2-1} \sech(\alpha[l-\mu]) 
  \sech(\alpha[l+1-\mu])}{\sum_{l=-L/2}^{L/2} \sech(\alpha[l-\mu]) 
  \sech(\alpha[l-\mu])}.
\end{eqnarray}

Let $P$ be the orthogonal projection to the subspace of perturbations 
considered so far, i.e.\ the span of $\psi^f$, where
$f(x) = \frac{1}{2} \sech(\alpha[l_x - \mu]) \phi(x)$.
Then the projection of $H$ to this subspace is $PHP$.
We have determined that 
$PHP \psi^f = \psi^g$ where $g$ has in place of $\Phi$
\be
\Psi = - \frac{\sech(\alpha)}{2 R^2} \cdot 
  \frac{\sum_{l=-L/2}^{L/2-1} \sech(\alpha[l-\mu]) 
  \sech(\alpha[l+1-\mu])}{\sum_{l=-L/2}^{L/2} \sech(\alpha[l-\mu]) 
  \sech(\alpha[l-\mu])} \nabla^2 \Phi.
\ee
(We write $\nabla^2$ for the Laplacian.
The symbol $\Delta$ is reserved for the anisotropy.)
We should note that it really is necessary to consider $PHP$ instead 
of $H$.
The reason for this is that
\be
\label{newdef:xi}
\xi_{xy} = \frac{-2 \cosh(\alpha[l_x-\mu]) w_x \otimes v_y 
+ 2 \cosh(\alpha[l_y-\mu]) v_x \otimes w_y
+ 2 \sinh(\alpha) w_x \otimes w_y}
{\sqrt{2 \cosh(\alpha[l_x-\mu]) \cdot
2 \cosh(\alpha[l_y-\mu]) \cdot
2 \cosh(\alpha)}} ,
\ee
which means that $H$ does not preserve the total number of $v_x$'s or
$w_x$'s.
Thus the perturbations we have considered (those with a single
$w_x$) do not form an invariant subspace of $H$.

\section{Error Terms}
We now come to the task of tying-up some loose ends,
in order that non-rigorous approximations can be replaced by 
rigorous bounds.
We start with a simple lemma.

\begin{lemma}
Let $\Gamma$ be a finite subset of a lattice $L$.
Let $\Omega$ be the Voronoi domain of $\Gamma$ with respect to
$L$, and let $\Omega_0$ be the Voronoi domain for the
single site $0 \in L$.
Then, for a smooth function $\phi : \Omega \to \C$,
\be
\Big| \frac{1}{|\Gamma|} \sum_{x \in \Gamma} u(x)
  - \frac{1}{m(\Omega)} \int_{\Omega} \phi(x)\, dx \Big| 
  < \|\partial^2 \phi\|_{op,\infty} \cdot \frac{1}{m(\Omega_0)} 
  \int_{\Omega_0} \frac{|x|^2}{2} dx ,
\ee
where $\partial^2 \phi$ is the second-derivative matrix and
\be
\|\partial^2 \phi\|_{op,\infty}
  = \sup_{x \in \Omega} \, \sup_{v \in \Rl^d \setminus 0} 
  \frac{v \cdot \partial^2 u(x) v}{v \cdot v} .
\ee
Note that the second moment $m(\Omega_0)^{-1} \int |x|^2\, dx$
is bounded by the radius of the Voronoi domain, which is in turn
bounded by the distance of nearest neighbors of $L$.
\end{lemma}

\begin{proof}
For the Voronoi domain $\Omega_0$ of $0$, we observe that
\begin{eqnarray*}
\frac{1}{m(\Omega_0)} \int_{\Omega_0} \phi(x)\, dx - \phi(0)
  &=& \frac{1}{m(\Omega_0)} \int_{\Omega_0} [\phi(x) - \phi(0)]\, dx \\
  &=& \frac{1}{m(\Omega_0)} \int_{\Omega_0} \int_0^1
    \nabla \phi(t x) \cdot x\, dt\, dx \\
  &=& \frac{1}{m(\Omega_0)} \int_{\Omega_0} \int_0^1 \int_0^t
  x \cdot \nabla^2 \phi(s x) x\, ds\, dt\, dx \\
  && + \frac{1}{m(\Omega_0)} \int_{\Omega_0} \nabla \phi(0) \cdot x\, dx\\
  &=& \frac{1}{m(\Omega_0)} \int_{\Omega_0} \int_0^1 (1-s)
  x \cdot \partial^2 \phi(s x) x\, ds\, dx \\
  && + \nabla \phi(0) \cdot \frac{1}{m(\Omega_0)} \int_{\Omega_0} x\, dx .
\end{eqnarray*}
But the centroid of $\Omega_0$ is 0.
Thus
\be
\label{dim}
\Big| \frac{1}{m(\Omega_0)} \int_{\Omega_0} \phi(x)\, dx - \phi(0) \Big|
  \leq \frac{1}{m(\Omega_0)} \int_{\Omega_0} \frac{|x|^2}{2} dx
  \times \|\partial^2 \phi\|_{op,\infty} .
\ee
The lemma follows by decomposing $\Omega$ into the $|\Gamma|$ affine copies
of $\Omega_0$, one for each site, and adding the inequalities obtained
from \eq{dim}.
\end{proof}
Using the result of this lemma, we make rigorous the approximation of
\eq{norm:approx}.
Thus
\be
\ip{\psi^f}{\psi^f}
  = |\Gamma| \ \sum_{l=-L/2}^{L/2} |F(l)|^2 \cdot
  \left(\frac{1}{m(\Omega)} \int_{\Omega} |\Phi(x)|^2\, dx + \epsilon_1
  \right),
\ee
where
\be
|\epsilon_1| \leq \frac{1}{R^2} 
  \|\partial^2 |\Phi|^2 \|_{op,\infty}.
\ee
(We have used the fact that the distance between nearest-neighbors
for $\Gamma$ is $\sqrt{2}$.)
In order to fix the approximation of \eq{energy:approx}, we begin
with the elementary bound
$|\phi(x+e_j) - \phi(x) - \nabla \phi(x) \cdot e_j| 
  < \frac{1}{2} \|\partial^2 \phi\|_{op,\infty}$
and its natural successor
\be
\Big| \sum_{j=1}^d |\phi(x+e_j) - \phi(x)|^2 
  - \|\nabla \phi(x)\|^2 \Big|
  < d \Big( \|\nabla \phi\|_{\infty}
  + \frac{1}{4} \|\partial^2 \phi\|_{op,\infty} \Big)
  \|\partial^2 \phi\|_{op,\infty} .
\ee
Using this estimate, as well as the lemma, we may replace 
\eq{energy:approx} with
\begin{eqnarray}
\label{energy:bound}
\op{\psi^f}{H}{\psi^f}
  &\approx& \frac{|\Gamma|}{8 R^2 \cosh(\alpha)} \sum_{l=-L/2}^{L/2} 
  \sech(\alpha[l-\mu]) \sec(\alpha[l+1-\mu]) \nonumber \\
  && \Bigg( \frac{1}{m(\Omega)} \int_\Omega |\nabla \Phi(x)|^2 dx 
  + \epsilon_2 + \epsilon_3 \Bigg),
\end{eqnarray}
where
\be 
|\epsilon_2| \leq \frac{d}{R} \Big(\|\nabla \Phi\|_{\infty}
  + \frac{1}{4 R} \|\partial^2 \Phi\|_{op,\infty} \Big)
  \|\partial^2 \Phi\|_{op,\infty} ,
\ee
and
\be
|\epsilon_3| \leq \frac{1}{R^2} 
  \|\partial^2 |\nabla \Phi|^2 \|_{op,\infty}.
\ee

\section*{Acknowledgements}

O.B. was supported by Fapesp under grant 97/14430-2. B.N. was partially
supported by the National Science Foundation under grant \# DMS-9706599.

\end{document}